\documentclass[useAMS,usenatbib]{mnras}
\usepackage[toc]{appendix}
\usepackage{graphicx}
\usepackage{verbatim} 
\usepackage{color} 
\input epsf
\newcommand{\Vvec}{\mbox{\bf V}}

\newcommand{\xvec}{\mbox{\bf x}}
\newcommand{\yvec}{\mbox{\bf y}}

\newcommand{\zvec}{\mbox{\bf z}}

\newcommand{\Omegavec}{\mbox{\boldmath $\Omega$}}

\title[The convective stability of baroclinic discs]{The convective stability of fully stratified baroclinic discs}
\author[Francesco Volponi]{Francesco Volponi\thanks{email: foxonif@yahoo.co.jp} 
\\{University of Information Science and Technology "St. Paul The Apostle", 6000 Ohrid, Macedonia}}

\begin{document}

\date{}
\maketitle
\begin{abstract}
We examine the convective stability of hydrodynamic discs with full stratification in the local approximation and in the presence of thermal diffusion (or relaxation). Various branches of the relevant axisymmetric dispersion relation derived by Urpin (2003) are discussed. We find that when the vertical Richardson number is larger than or equal to the radial one  (i.e. $|Ri_z|\geq|Ri_x|$) and wavenumbers are comparable (i.e. $|k_x|\sim|k_z|$) the disc becomes unstable, even in the presence of radial and vertical stratifications with $Ri_x>0$ and $Ri_z>0$. The origin of this resides in an hybrid radial-vertical Richardson number. We propose an equilibrium profile with temperature depending on the radial and vertical coordinates and with $Ri_z>0$ for which this destabilization mechanism occurs. We notice as well that the dispersion relation of the "convective overstability" is the branch of the one here discussed in the limit $|k_z|\gg|k_x|$ (i. e. two-dimensional disc).
\end{abstract}
\clearpage
\section{Introduction}
Astrophysical discs composed of ionized gases accrete toward a central object because subject to the magnetorotational instability (MRI) (Balbus \& Hawley 1991), which induces transport of angular momentum outwardly.

Extended zones of protoplanetary discs are, however, scarcely ionized (Gammie 1996). In these regions the MRI drive is absent and other instabilities of hydrodynamical nature must be active in order to explain the process of planet formation.

Among various candidates the strongest are the vertical shear instability (VSI) and the subcritical baroclinic instability (SBI).

The VSI is a linear process studied first in differentially rotating stars (Goldreich \& Schubert 1967; Fricke 1968) and then investigated in discs linearly by Urpin (2003) and nonlinearly by Nelson, Gressel \& Umurhan  (2013). Urpin's analysis investigates quite generally the local stability of a fully stratified and thermally diffusing disc in the presence not only of vertical shear but also of convection (radial and vertical). Its treatment provides a unified axisymmetric dispersion relation for all the aforementioned processes.

The SBI (Klahr \& Bodenheimer 2003; Lesur \& Papaloizou 2010) is of non-linear nature and was recently related to a linear growth mechanism (Klahr \& Hubbard 2014; Lyra 2014), denominated convective overstability, capable of amplifying small disturbances to finite perturbations, the seeds from which the SBI developes. Convective overstability is a radial convective instability arising in the presence of thermal relaxation (or thermal diffusion) when the radial Richardson number, $Ri_x$, is negative and for vertical wavelengths much shorter than the radial ones. Here we will first discuss the convective overstability in the context of the general theory of Urpin (2003). We will show that the dispersion relation of Klahr \& Hubbard (2014)  is the branch of the dispersion relation of Urpin (2003) in the regime $k_z \gg k_x$ ($k_z$ and $k_x$ are the vertical and radial wavenumbers respectively). As well the growth rates are essentially the same. The difference between the thermally diffusing model of Urpin (2003) and the thermally relaxing model of Klahr \& Hubbard (2014) stays in the fact that in the latter the growth rates are independent of the perturbations wavelengths, whereas in the first one perturbations with intermediate wavelengths grow fastest.

The main focus of this paper is the investigation of the branch of the dispersion relation of Urpin (2003) corresponding to $|Ri_z| \geq |Ri_x|$ and $|k_x| \sim |k_z|$. We will show that in this sector, even for radial and vertical stratifications with $Ri_x>0$ and $Ri_z>0$, the disc can become unstable when the ratio $|{{L_S}_x}/{{L_S}_z}|$ is significantly larger or smaller than $1$ (${{L_S}_x}$ and ${{L_S}_z}$ are the radial and vertical entropy lenght scales respectively). The origin of this slightly counterintuitive behaviour stays in the hybrid Richardson number $Ri_{xz}={\rm sgn}(k_x k_z)[({{L_S}_x}/{{L_S}_z})Ri_x+({{L_S}_z}/{{L_S}_x})Ri_z]$.

In order to provide a tangible realization of this instability and to connect these linear results to recent nonlinear simulations we present an equilibrium, with temperature profile function of both radial and vertical coordinates, for which the disc has vertical stratification with $Ri_z>0$ and where the destabilization mechanism here described occurs. For such equilibria we determine the contours in the $(R,z)$ plane where the instability developes. The growth rates, which in general vary greatly along the contours, are as well determined at some representative locations both in the case of thermal diffusion and of thermal relaxation.

The configuration considered generalizes the equilibria studied in Nelson, Gressel and Umurhan (2013), which were either vertically isothermal (i.e. $T=T(R)$, where $T$ denotes temperature) or vertically isentropic (i.e. $Ri_z=0$), to an equilibrium where  $T=T(R,z)$ and $Ri_z \neq 0$. Our equilibrium should, however, be viewed more as a proof of concept for the instability mechanism here presented than as an attempt to a detailed description of a protoplanetary disc.
\section{Axisymmetric baroclinic disc}
The local stability of a baroclinic disc with full stratification and in the presence of thermal diffusion was studied in full generality by Urpin (2003). We reformulate here his analysis along the lines of Volponi (2014). We start from the shearing sheet equations
\begin{equation}
\partial_{t}{\rho}+\nabla \cdot{\rho \Vvec}=0,
\label{eq1.1}
\end{equation}
\begin{equation}
\partial_{t}{\Vvec}+\Vvec\cdot\nabla{\Vvec}=-\frac{\nabla{P}}{\rho}-2{\Omegavec}\times{\Vvec}+2q{\Omega}^2 x \hat{{\xvec}}-{\Omega}^2 z \hat{{\zvec}},
\label{eq1.2}
\end{equation}
\begin{equation}
\partial_{t}{(\ln{S})}+\Vvec\cdot\nabla{(\ln{S})}={\chi}_d \Delta (\ln{\frac{T}{T_{\rm e}}}),
\label{eq1.3}
\end{equation}
where $\rho$, $P$ and ${T}$ are density, pressure and temperature, $T_{\rm e}$ is the equilibrium temperature, $\Vvec$ is the fluid velocity, $S=P{\rho}^{-\gamma}$ is a measure of the fluid entropy, $\gamma$ is the adiabatic index, $\Omega$ is the local rotation frequency, $q$ is the shear parameter ($q=1.5$ for Keplerian rotation) and ${\chi}_d$ is the thermal diffusion coefficient. The term $-2{\Omegavec}\times{\Vvec}$ is the Coriolis term, $2q{\Omega}^2 x \hat{{\xvec}}$ is the tidal expansion of the effective potential and $-{\Omega}^2 z \hat{{\zvec}}\equiv{-g_z}\hat{{\zvec}}$ is the vertical gravitational acceleration. The equations are expressed in terms of the pseudo-Cartesian coordinates $x=R-R^*$, $y=R^*(\phi-{\phi}^*)$ and $z$  ($R^*$ and ${\phi}^*$ are reference radius and angle).\\
The disc consists of an ideal gas with equation of state
\begin{equation}
P=\frac{\cal{R}}{\mu}\rho T,
\label{eq1.4}
\end{equation}
where $\cal{R}$ and $\mu$ are the gas constant and molecular weight respectively.\\
The equilibrium is constrained by the equations
\begin{equation}
\frac{\partial_{z}{P_{\rm e}}}{\rho_{\rm e}}=-{\Omega}^2 z.
\label{eq1.5A}
\end{equation}
and
\begin{equation}
{\Vvec}_{\rm e}(x, z)=\Big[-q\Omega x+\frac{\partial_{x}P_{\rm e}(x,z)}{2 \Omega {\rho}_{\rm e}(x,z)}\Big]\hat{\yvec},
\label{eq1.5B}
\end{equation}
which provide vertical structure and velocity field respectively. The subscript "${\rm e}$" denotes equilibrium quantities.

Deriving equation~(\ref{eq1.5A}) with respect to $x$, equation~(\ref{eq1.5B}) with respect to $z$ and introducing the sound speed ${c_s}^2=\gamma P_{\rm e} /\rho_{\rm e}$, we obtain the following expression for the vertical velocity shear
\begin{eqnarray}
\partial_{z}{V}_{\rm e}(x, z)\hspace{-0.3cm}&=&\hspace{-0.25cm}\partial_{z}\Big[\frac{\partial_{x}P_{\rm e}(x,z)}{2 \Omega {\rho}_{\rm e}(x,z)} \Big] 
\label{eq1.5C}
\\ &=& \hspace{-0.25cm} \frac{1}{2\Omega}\Big[\frac{\partial_{z}c_s^2}{\gamma}\Big(\frac{\partial_{x}\rho_{\rm e}}{\rho_{\rm e}}+\frac{\partial_{x}c_s^2}{c_s^2}\Big)+\frac{\partial_{x}c_s^2}{c_s^2}{{\Omega}^2 z}\Big] \nonumber.
\end{eqnarray}

For discs with equilibrium temperature profiles smooth and symmetric about the midplane (i. e. depending on $z^2$ rather than $z$) ${V}_{\rm e}$ becomes a function of $z^2$. This is due to the linear dependence of the vertical gravity on $z$. In atmospheres, instead, which are modelled with a constant $g_z$ (Goldreich \& Schubert 1967), ${V}_{\rm e}$ is linear in $z$. 

In this study, by parametrizing the vertical shear with a constant coefficient $\bar{A}_z$, we consider velocity equilibria of the type (Volponi 2014)
\begin{equation}
{\Vvec}_{\rm e}(x, z)=\Big[-q\Omega x+\frac{\partial_{x}P_{\rm e}(x,0)}{2 \Omega {\rho}_{\rm e}(x,0)}+\bar{A}_z z\Big]\hat{\yvec}.
\label{eq1.6}
\end{equation}
Equation~(\ref{eq1.6}) is valid overall in atmospheres, while in discs holds at finite altitude, since in the neighborhoods of the midplane $\bar{A}_z=0$.

Linearizing equations~(\ref{eq1.1})-(\ref{eq1.3}) about the equilibria discussed above,
short wavelength axisymmetric Eulerian perturbations of the type
\begin{equation}
{\delta}'(t,x,y,z)=\hat{{\delta}'}(t)e^{i{K}_x x+i{K}_z z}
\label{eq1.7}
\end{equation}
 evolve according to the equations
\begin{equation}
\partial_{t}\frac{\hat{\rho'}}{\rho_{\rm e}}+\frac{\hat{v'_x}}{{L_{\rho}}_x}+\frac{\hat{v'_z}}{{L_{\rho}}_z}+i{K}_x\hat{v'_x}+i{K}_z\hat{v'_z}=0,
\label{eq1.8}
\end{equation}
\begin{equation}
\partial_{t}\hat{v'_x}= 2\Omega \hat{v'_y}-i{K}_x \frac{\hat{P'}}{{\rho}_{\rm e}}+\frac{c_s^2}{{L_P}_x}\frac{\hat{\rho'}}{{\rho}_{\rm e}},
\label{eq1.9}
\end{equation}
\begin{equation}
\partial_{t}\hat{v'_y}=-(2-\tilde{q})\Omega \hat{v'_x}-\bar{A}_z\hat{v'_z},
\label{eq1.10}
\end{equation}
\begin{equation}
\partial_{t}\hat{v'_z}=-i\tilde{K}_z\frac{\hat{P'}}{{\rho}_{\rm e}}+\frac{c_s^2}{{L_P}_z}\frac{\hat{\rho'}}{{\rho}_{\rm e}},
\label{eq1.11}
\end{equation}
\begin{equation}
\partial_{t}\Big(\frac{\hat{P'}}{P_{\rm e}}-\gamma\frac{\hat{\rho'}}{\rho_{\rm e}}\Big)+\gamma\frac{\hat{v'_x}}{{L_S}_x}+\gamma\frac{\hat{v'_z}}{{L_S}_z}=-{\chi}_d K^2\Big(\frac{\hat{P'}}{P_{\rm e}}-\frac{\hat{\rho'}}{\rho_{\rm e}}\Big),
\label{eq1.12}
\end{equation}
where $K^2=K_x^2+K_z^2$ and $\tilde{q}(x){\Omega}=-\frac{d{V}_{\rm e}(x)}{dx}$ is an effective shear rate (Johnson and Gammie 2005) varying with $x$.

The radial and vertical length scales for pressure, density and entropy are defined by
\begin{equation}
\frac{1}{{L_P}_x}\equiv\frac{\partial_{x}P_{\rm e}}{\gamma P_{\rm e}}=\frac{1}{{L_{\rho}}_x}+\frac{1}{{L_S}_x}\equiv\frac{\partial_{x}{\rho}_{\rm e}}{ {\rho}_{\rm e}}+\frac{\partial_{x}S_{\rm e}}{\gamma S_{\rm e}},
\label{eq1.18}
\end{equation}
\begin{equation}
\frac{1}{{L_P}_z}\equiv\frac{\partial_{z}P_{\rm e}}{\gamma P_{\rm e}}=\frac{1}{{L_{\rho}}_z}+\frac{1}{{L_S}_z}\equiv\frac{\partial_{z}{\rho}_{\rm e}}{ {\rho}_{\rm e}}+\frac{\partial_{z}S_{\rm e}}{\gamma S_{\rm e}}.
\label{eq1.19}
\end{equation}

We are interested in the evolution of incompressive perturbations and we work therefore in the Boussinesq approximation, which consists in transforming equation~(\ref{eq1.8}) into an incompressibility condition and in neglecting the pressure terms in equation~(\ref{eq1.12}). Equations~(\ref{eq1.8}) and~(\ref{eq1.12}) become respectively
\begin{equation}
{K}_x\hat{v'_x}+{K}_z \hat{v'_z}=0
\label{eq1.8b}
\end{equation}
and
\begin{equation}
\partial_{t}\frac{\hat{\rho'}}{\rho_{\rm e}}=\frac{\hat{v'_x}}{{L_S}_x}+\frac{\hat{v'_z}}{{L_S}_z}-\frac{{\chi}_d}{\gamma} K^2 \frac{\hat{\rho'}}{\rho_{\rm e}}.
\label{eq1.12b}
\end{equation}
Deriving with respect to time equation~(\ref{eq1.8b}) and then in the expression obtained substituting~(\ref{eq1.9}) and (\ref{eq1.11})
we have
\begin{eqnarray}
i\frac{\hat{P'}}{{\rho}_{\rm e}}=\frac{1}{{K}^2}\Big[\Big({K}_x\frac{c_s^2}{{L_P}_x}+{K}_z\frac{c_s^2}{{L_P}_z} \Big)\frac{\hat{\rho'}}{\rho_{\rm e}} + 2\Omega {K}_x \hat{v'_y}\Big].
\label{eq1.20}
\end{eqnarray}
By means of equation~(\ref{eq1.20}) equations~(\ref{eq1.9})-(\ref{eq1.11}) become 
\begin{eqnarray}
\partial_{t}\hat{v'_x}=  2\Omega  \Big(1-\frac{{{K}_x^2}}{{{K}^2}}\Big)\hat{v'_y}+\frac{c_s^2}{{L_P}_x}\Big(1-\frac{{{K}_x^2}}{{{K}^2}}\Big) \frac{\hat{\rho'}}{{\rho}_{\rm e}}\nonumber \\-\frac{c_s^2}{{L_P}_z}\frac{{K}_z {K}_x }{{{K}^2}} \frac{\hat{\rho'}}{{\rho}_{\rm e}},
\label{eq1.21}
\end{eqnarray}
\begin{eqnarray}
\partial_{t}\hat{v'_y}= (\tilde{q}-2)\Omega\hat{v'_x}-\bar{A}_z\hat{v'_z},
\label{eq1.22}
\end{eqnarray}
\begin{eqnarray}
\partial_{t}\hat{v'_z}= - 2\Omega  \frac{{{K}_x{K}_z}}{{{K}^2}}\hat{v'_y} +\frac{c_s^2}{{L_P}_z}\Big(1-\frac{{{K}_z^2}}{{{K}^2}}\Big) \frac{\hat{\rho'}}{{\rho}_{\rm e}} \nonumber \\-\frac{c_s^2}{{L_P}_x}\frac{ {K}_z {K}_x }{{{K}^2}} \frac{\hat{\rho'}}{{\rho}_{\rm e}},
\label{eq1.23}
\end{eqnarray}
Normalizing time with ${\Omega}^{-1}$, velocities with ${L_S}_z \Omega$ and density with $\rho_{\rm e}$, we obtain for the evolution of the non-dimensional variables $v_x$, $v_y$, $v_z$ and $\rho$ the system
\begin{equation}
\partial_{t}{v_x}=  2 \frac{{{k}_z^2}}{{{k}^2}}{v_y} - \frac{{L_S}_x}{{L_S}_z}Ri_x\frac{{{k}_z^2}}{{{k}^2}}{{\rho}}+Ri_z\frac{ k_z {k}_x }{{{k}^2}} {{\rho}},
\label{eq1.31}
\end{equation}
\begin{equation}
\partial_{t}{v_y}= (\tilde{q}-2){v_x}-{A_z}{v_z},
\label{eq1.32}
\end{equation}
\begin{equation}
\partial_{t}{v_z}= - 2  \frac{{{k}_x{k}_z}}{{{k}^2}}{v_y} -Ri_z\frac{{{k}_x^2}}{{{k}^2}} {{\rho}}+\frac{{L_S}_x}{{L_S}_z}Ri_x\frac{ {k}_z {k}_x }{{{k}^2}} {{\rho}},
\label{eq1.33}
\end{equation}
\begin{equation}
\partial_{t}{{\rho}}=\frac{{L_S}_z}{{L_S}_x}v_x+{v_z}-{{k}^2}\frac{1}{Pe}{{\rho}},
\label{eq1.12bb}
\end{equation}
where $(k_x,k_z)\equiv{{L_S}_z}(K_x,K_z)$, and ${{{k}^2}}\equiv{{L_{S_z}^2}}{{{K}^2}}$ and $Pe={{{L_S^2}_z}\Omega \gamma}/{\chi}_d$ is the Peclet number. We introduced, as well, $A_z\equiv{\bar{A}_z}/{\Omega}$ and the Richardson numbers
\begin{equation}
Ri_x\equiv\frac{N_x^2}{{\Omega}^2}, \hspace{2cm} Ri_z\equiv\frac{N_z^2}{{\Omega}^2}.
\label{eq1.29}
\end{equation}
$N_x$ and $N_z$ are the Brunt-V\"{a}is\"{a}l\"{a} frequencies
\begin{equation}
N_x^2\equiv-\frac{c_s^2}{{L_S}_x{L_P}_x},  \hspace{0.5cm} N_z^2\equiv\frac{g_z}{{L_S}_z}=-\frac{c_s^2}{{L_S}_z{L_P}_z}.
\label{eq1.30}
\end{equation}
Assuming an exponential time dependence of the type $e^{st}$ for the perturbations, equations~(\ref{eq1.31})-(\ref{eq1.12bb}) lead to the following dispersion relation (Urpin 2003; Volponi 2014)
\begin{eqnarray}
s^3+k^2 Pe^{-1}s^2+s\bigg[\frac{k_z^2}{k^2}[2(2-\tilde{q})+Ri_x]- \nonumber \\  \frac{k_x k_z}{k^2}\Big(2{A_z}+\frac{{L_S}_x}{{L_S}_z}Ri_x+  \frac{{L_S}_z}{{L_S}_x}Ri_z\Big)+\frac{Ri_z k_x^2}{k^2}\bigg]+  \nonumber \\  Pe^{-1}[2k_z^2(2-\tilde{q})-{2k_x k_z}{A_z}]=0,
\label{eq1.35C}
\end{eqnarray}
Thermally relaxed models (Klahr \& Hubbard 2014) are essentially equivalent to the above formulation, the only difference being the replacement of the Laplacian operator ${\chi}_d \Delta$ in equation~(\ref{eq1.3}) with a multiplicative constant $-\chi_r$ where $\chi_r$ represents the inverse of the thermal time. This is tantamount to the substitution $K^2\chi_d \leftrightarrow \chi_r$  in equation~(\ref{eq1.12b}) (Urpin 2003) or $k^2 Pe^{-1} \leftrightarrow {\chi}_r/\gamma \Omega$ in terms of the non-dimensional quantities in equation~(\ref{eq1.35C}). We notice that in Klahr \& Hubbard (2014) ${\chi}_r$ is denoted as ${1}/{ \tau}$.

By casting equation~(\ref{eq1.35C}) in the form
\begin{equation}
s^3+a_2s^2+a_1s+a_0=0,
\label{eq1.35D}
\end{equation}
where
\begin{eqnarray}
a_2=k^2 Pe^{-1}, \nonumber \\ a_1=\bigg[\frac{k_z^2}{k^2}[2(2-\tilde{q})+Ri_x]-\frac{k_x k_z}{k^2}\Big(2{A_z}+\nonumber \\  \frac{{L_S}_x}{{L_S}_z}Ri_x+\frac{{L_S}_z}{{L_S}_x}Ri_z\Big)+\frac{Ri_z k_x^2}{k^2}\bigg], \nonumber\\
a_0=Pe^{-1}[2k_z^2(2-\tilde{q})-{2k_x k_z}{A_z}],
\label{eq1.35E}
\end{eqnarray}
the instability conditions are (Urpin 2003)
\begin{equation}
a_0<0, \hspace{1cm} a_1 a_2<a_0, \hspace{1cm} a_2<0.
\label{eq1.35F}
\end{equation}

The first and the second of the above inequalities read
\begin{equation}
Pe^{-1}[2k_z^2(2-\tilde{q})-{2k_x k_z}{A_z}]<0
\label{eq1.35instvert}
\end{equation}
and 
\begin{equation}
\bigg[\frac{k_z^2}{k^2}Ri_x-\frac{k_x k_z}{k^2}\Big(\frac{{L_S}_x}{{L_S}_z}Ri_x+\frac{{L_S}_z}{{L_S}_x}Ri_z\Big)+\frac{Ri_z k_x^2}{k^2}\bigg]<0
\label{eq1.35instconv}
\end{equation}
respectively.
From~(\ref{eq1.35instvert}) it can be seen that for $|k_x| \gg |k_z|$ the discs is subject to the vertical shear instability when $A_z$ and $k_xk_z$ have the same sign, whereas~(\ref{eq1.35instconv}) concerns the convective stability of the disc. Volponi (2014) discussed the combined effect of the vertical shear and vertical convective instabilities finding that the resulting growths are of mixed type in the sense that the growth rate is given by convection whereas the sign of the angular momentum transport is determined by the vertical shear.
Here we would like to concentrate on the second of the above conditions and discuss more in detail the convective stability of the disc in different regimes characterized by the relative strength of $Ri_x$ and $Ri_z$. For each of these regimes we will discuss the limits $|k_z| \gg |k_x|$, $|k_x| \gg |k_z|$ and $|k_x|\sim |k_z|$.
\subsection{Regime A: $|Ri_x| \gg |Ri_z|$}
Here we consider stratifications which are stronger radially than vertically.\\\\
$|k_z| \gg |k_x|$: In this limit we are essentially dealing with a two-dimensional disc (i.e. vertical structure is neglected). The dispersion relation reads
\begin{eqnarray}
s^3+k^2 Pe^{-1}s^2+s\frac{k_z^2}{k^2}[2(2-\tilde{q})+Ri_x]+  \nonumber \\  Pe^{-1}[2k_z^2(2-\tilde{q})]=0,
\label{eq1.35CC}
\end{eqnarray}
which is the same relation obtained by Klahr and Hubbard (2014), considering that $k^2 Pe^{-1}$ corresponds to their $1/\gamma\tau\Omega$, $k_z \simeq k$ for $k_z \gg k_x$ and that there time dependence of perturbations was assumed to be of the form $e^{-i\omega t}$. The instability condition~(\ref{eq1.35instconv}) becomes simply $Ri_x<0$.\\\\
$|k_x| \gg |k_z|$: In this case equation~(\ref{eq1.35instconv}) becomes
\begin{equation}
\bigg[\frac{k_z^2}{k^2}Ri_x-\frac{k_x k_z}{k^2}\frac{{L_S}_x}{{L_S}_z}Ri_x+\frac{Ri_z k_x^2}{k^2}\bigg]<0.
\label{eq1.35instconv2}
\end{equation}
and a variety of different evolutions are possible in principle depending on the specific values of $Ri_x$, $Ri_z$, $k_x$ and $k_z$. However, if $A_z$ and the product $k_x k_z$ have the same sign, the vertical shear instability dominates.\\\\
$|k_x| \sim |k_z|$:
The instability condition becomes
\begin{equation}
Ri_x-{\rm sgn}(k_x k_z)\Big(\frac{{L_S}_x}{{L_S}_z}Ri_x+\frac{{L_S}_z}{{L_S}_x}Ri_z\Big)<0,
\label{eq1.35instconv3}
\end{equation}
which can be alternatively expressed as
\begin{equation}
Ri_x\Big[1-{\rm sgn}(k_x k_z)\Big(\frac{{L_S}_x}{{L_S}_z}+\frac{{L_P}_x}{{L_P}_z}\Big)\Big]<0.
\label{eq1.35instconv3a}
\end{equation}
Characteristic values of $Ri_x$, $Ri_z$ and ${{L_S}_x}/{{L_S}_z}$ in the disc domain can vary greatly with the equilibrium considered. It is therefore difficult to discuss equation~(\ref{eq1.35instconv3}) without reference to a specific configuration. If $|({{L_S}_x}/{{L_S}_z})Ri_x|\gg|({{L_S}_z}/{{L_S}_x})Ri_z|$ (i.e. $|{{L_P}_x}/{{L_P}_z}|\ll|{{L_S}_x}/{{L_S}_z}|$), however,~(\ref{eq1.35instconv3}) simplifies to
\begin{equation}
Ri_x\Big(1-{\rm sgn}(k_x k_z)\frac{{L_S}_x}{{L_S}_z}\Big)<0.
\label{eq1.35instconv3b}
\end{equation}
In the case of  $k_x k_z>0$ instability ensues for $Ri_x<0$  when ${{L_S}_x}/{{L_S}_z}<1$ and for $Ri_x>0$ when ${{L_S}_x}/{{L_S}_z}>1$. \\
If $k_x k_z<0$, instead, instability occurs for ${{L_S}_x}/{{L_S}_z}<-1$ when $Ri_x>0$ and for ${{L_S}_x}/{{L_S}_z}>-1$ when $Ri_x<0$.\\
If $|({{L_S}_x}/{{L_S}_z})Ri_x|\ll|({{L_S}_z}/{{L_S}_x})Ri_z|$ (i.e. $|{{L_P}_x}/{{L_P}_z}|\gg|{{L_S}_x}/{{L_S}_z}|$) instead, equation~(\ref{eq1.35instconv3}) becomes
\begin{equation}
Ri_x\Big(1-{\rm sgn}(k_x k_z)\frac{{L_P}_x}{{L_P}_z}\Big)<0.
\label{eq1.35instconv3c}
\end{equation}
In~(\ref{eq1.35instconv3c}) ${{L_P}_x}/{{L_P}_z}$ has the same role that ${{L_S}_x}/{{L_S}_z}$ has in equation~(\ref{eq1.35instconv3b}) and therefore
observations analogous to the ones above given pertain.
\subsection{Regime B: $|Ri_z| \gg |Ri_x|$}
Here we focus on stratifications which are stronger vertically than radially.\\\\
$|k_z| \gg |k_x|$: For such perturbations the epiciclic frequency dominates the vertical shear (see equation~(\ref{eq1.35instvert})) and therefore the disc is unstable if
\begin{equation}
\bigg[\frac{k_z^2}{k^2}Ri_x-\frac{k_x k_z}{k^2}\frac{{L_S}_z}{{L_S}_x}Ri_z+\frac{Ri_z k_x^2}{k^2}\bigg]<0.
\label{eq1.35instconv4}
\end{equation}
Various types of evolutions are possible depending on the specific value of wavenumbers and Richardson numbers.\\\\
$|k_x| \gg |k_z|$: Equation~(\ref{eq1.35instconv}) becomes simply $Ri_z<0$. If $A_z$ and the product $k_x k_z$ have the same sign, however, the vertical shear instability drive is present. This case was discussed in detail in Volponi (2014). For stable stratification (i.e. $Ri_z>0$) we expect an evolution dominated by the vertical shear instability. For unstable stratification instead (i.e. $Ri_z<0$) we expect evolutions of mixed types where growth rate is the convective one and the sign of transport is determined by the vertical shear.\\\\
$|k_x| \sim |k_z|$:
Equation~(\ref{eq1.35instconv}) reads
\begin{equation}
Ri_z-{\rm sgn}(k_x k_z)\Big(\frac{{L_S}_x}{{L_S}_z}Ri_x+\frac{{L_S}_z}{{L_S}_x}Ri_z\Big)<0,
\label{eq1.35instconv5}
\end{equation}
which can be as well cast  in the form
\begin{equation}
Ri_z\Big[1-{\rm sgn}(k_x k_z)\Big(\frac{{L_P}_z}{{L_P}_x}+\frac{{L_S}_z}{{L_S}_x}\Big)\Big]<0.
\label{eq1.35instconv5b}
\end{equation}
This case is formally analogous to the corresponding one discussed in the previous section but with the $x$ and $z$ coordinates interchanged. As mentioned there, any discussion of the above condition strongly depends on the particular equilibrium considered. We will come back in greater detail to equation~(\ref{eq1.35instconv5}) in section 3.\\
If $|({{L_S}_x}/{{L_S}_z})Ri_x|\ll|({{L_S}_z}/{{L_S}_x})Ri_z|$ (i.e. $|{{L_P}_z}/{{L_P}_x}|\ll|{{L_S}_z}/{{L_S}_x}|$)~(\ref{eq1.35instconv5}) reduces to
\begin{equation}
Ri_z\Big(1-{\rm sgn}(k_x k_z)\frac{{L_S}_z}{{L_S}_x}\Big)<0.
\label{eq1.35instconv6}
\end{equation}
In case of positive $k_x k_z$, for $Ri_z<0$ the disc is unstable when ${{L_S}_z}/{{L_S}_x}<1$ and for $Ri_z>0$ when ${{L_S}_z}/{{L_S}_x}>1$. \\
If $k_x k_z<0$, instead, instability occurs for ${{L_S}_z}/{{L_S}_x}<-1$ when $Ri_z>0$ and for ${{L_S}_z}/{{L_S}_x}>-1$ when $Ri_z<0$.\\
If $|({{L_S}_x}/{{L_S}_z})Ri_x|\gg|({{L_S}_z}/{{L_S}_x})Ri_z|$ (i.e. $|{{L_P}_z}/{{L_P}_x}|\gg|{{L_S}_z}/{{L_S}_x}|$)~(\ref{eq1.35instconv5}) becomes
\begin{equation}
Ri_z\Big(1-{\rm sgn}(k_x k_z)\frac{{L_P}_z}{{L_P}_x}\Big)<0.
\label{eq1.35instconv6b}
\end{equation}
Everything stated above about equation~(\ref{eq1.35instconv6}) holds as well for equation~(\ref{eq1.35instconv6b}) when substituting ${{L_S}_z}/{{L_S}_x}$ with ${{L_P}_z}/{{L_P}_x}$.
\subsection{Regime C: $|Ri_z| \sim |Ri_x|$}
Here we consider radial and vertical stratifications which are comparable.\\\\
$|k_z| \gg |k_x|$: The instability condition~(\ref{eq1.35instconv}) assumes the simple form $Ri_x<0$. We are dealing with a two-dimensional disc and the same remarks made for the corresponding case of subsection 2.1 pertain.\\\\
$|k_x| \gg |k_z|$: Equation~(\ref{eq1.35instconv}) becomes simply $Ri_z<0$. In this case, though, the disc is as well vertical shear unstable. Therefore if $Ri_z>0$ the vertical shear instability dominates the evolution. We observed as well that the larger $Ri_z$ the weaker the vertical shear growth rate. If $Ri_z<0$ a mixed type of evolution of the type described in Volponi (2014) follows.\\\\
$|k_z| \sim |k_x|$: This is the most interesting subcase for the present study along with the corresponding case of section 2.2.  Equation~(\ref{eq1.35instconv}) reads:
\begin{equation}
Ri_x\Big(1-{\rm sgn}(k_x k_z)\frac{{L_S}_x}{{L_S}_z}\Big)+Ri_z\Big(1-{\rm sgn}(k_x k_z)\frac{{L_S}_z}{{L_S}_x}\Big)<0.
\label{eq1.35instconv8}
\end{equation}
Let's discuss in detail the case $k_xk_z>0$. When $Ri_x>0$ and $Ri_z>0$ the disc is unstable if ${{L_S}_x}/{{L_S}_z}>0$ (excluding the case ${{L_S}_x}/{{L_S}_z}=1$) and stable otherwise. This is a situation which is possible in a real disc. It is slightly counterintuitive how a disc with positive radial and vertical Richardson numbers can become convectively unstable. However from equation~(\ref{eq1.35instconv8}) it can be noticed that the critical term driving the disc to instability is the hybrid Richardson number
\begin{equation}
Ri_{xz}={\rm sgn}(k_x k_z)\Big(\frac{{L_S}_x}{{L_S}_z}Ri_x+\frac{{L_S}_z}{{L_S}_x}Ri_z\Big).
\label{eq1.35HybRichxz}
\end{equation}
When discs are fully stratified, instability drives arise not only from purely radial and vertical gradients but as well from mixed radial-vertical ones.\\\\
When $Ri_x<0$ and $Ri_z<0$ the disc is unstable if ${{L_S}_x}/{{L_S}_z}<0$ and stable otherwise.\\\\
When $Ri_x>0$ and $Ri_z<0$ we have instability for ${{L_S}_x}/{{L_S}_z}>1$ or  $-1<{{L_S}_x}/{{L_S}_z}<0$.\\\\
When $Ri_x<0$ and $Ri_z>0$ instability occurs for ${{L_S}_x}/{{L_S}_z}<-1$ or  $0<{{L_S}_x}/{{L_S}_z}<1$.\\\\
If $k_x \sim -k_z$ considerations similar to the ones developed above pertain.\\
The bottom line of the above classification is that if $|k_x|$ and $|k_z|$ are comparable the disc is potentially unstable even when $Ri_x$ and $Ri_z$ are positive. These instabilities are just different sectors of equation~(\ref{eq1.35instconv}).
\subsection{Growth rate} 
In this section we revisit what found by Urpin (2003) concerning the growth rate of perturbations. This will allow us a close comparison with growth rates obtained by Klahr \& Hubbard (2014).\\
First of all we simplify the notation. By defining 
\begin{eqnarray}
\eta \equiv Pe^{-1},  \hspace{1cm} B^2 \equiv 2(2-\tilde{q}),\nonumber \\
C \equiv \frac{k_z^2}{k^2}(B^2+Ri_x)- \frac{k_x k_z}{k^2}\Big(2{A_z}+\frac{ Ri_{xz}}{{\rm sgn}(k_x k_z)}\Big)+ \nonumber \\\frac{Ri_z k_x^2}{k^2},\nonumber \\
D \equiv k_z^2 B^2-{2k_x k_z}{A_z},
\end{eqnarray}
equation~(\ref{eq1.35C}) can be cast in the form
\begin{eqnarray}
s^3+k^2 \eta s^2+Cs+\eta D=0.
\label{eq1.35Csimplified}
\end{eqnarray}
With $s=r+i\omega$ equation~(\ref{eq1.35Csimplified}) splits in two relations one for its real part and the other for its imaginary part:
\begin{eqnarray}
r^3-3r{\omega}^2+k^2 \eta (r^2-{\omega}^2)+rC+\eta D=0
\label{eq1.35CsimplifiedReal}
\end{eqnarray}
and 
\begin{eqnarray}
{\omega}^2=3r^2+k^2 \eta 2r+ C.
\label{eq1.35CsimplifiedImaginary}
\end{eqnarray}
For $\omega\neq0$ we can substitute~(\ref{eq1.35CsimplifiedImaginary}) in~(\ref{eq1.35CsimplifiedReal}) obtaining
\begin{eqnarray}
2r(2r+k^2 \eta)^2+2rC+\eta E=0,
\label{eq1.35Csimplified2}
\end{eqnarray}
where
\begin{eqnarray}
E \equiv k^2C-D=Ri_x {k_z}^2-k_x k_z \frac{ Ri_{xz}}{{\rm sgn}(k_x k_z)}+Ri_z{k_x}^2.
\label{eq1.35CsimplifiedE}
\end{eqnarray}
In the limit $r\ll k^2 \eta$ the growth rate $r$ is
\begin{eqnarray}
r_d=-\frac{1}{2}\frac{\eta E}{k^4 {\eta}^2+C}.
\label{eq1.35CsimplifiedGrowthratediff}
\end{eqnarray}

We conclude this subsection noting that by using the prescription $k^2 Pe^{-1} \leftrightarrow {\chi}_r/\gamma \Omega$ we can obtain from equation~(\ref{eq1.35CsimplifiedGrowthratediff}) the growth rate pertaining to the thermally relaxed case as
\begin{eqnarray}
r_r=-\frac{1}{2k^2}\frac{\gamma{\bar{\chi}}_r E}{ {{\bar{\chi}}_r}^2+{\gamma}^2C},
\label{eq1.35CsimplifiedGrowthraterelax}
\end{eqnarray}
where ${\bar{\chi}}_r={\chi}_r/\Omega$. Equation~(\ref{eq1.35CsimplifiedGrowthraterelax}) is just a reformulation of the growth rate derived by Urpin (2003). In Urpin's analysis the growth rate for the cases ${ {{\bar{\chi}}_r}^2\ll |C|}$ and ${ {{\bar{\chi}}_r}^2\gg |C|}$ was respectively given in equations (30) and (32) of that paper, where ${\chi}_r$ was denoted as ${\omega}_{\chi}$. We notice that, in the limit $|k_z|\gg |k_x|$, equation~(\ref{eq1.35CsimplifiedGrowthraterelax}) is identical to equation (27) of Klahr \& Hubbard (2014).\\
$r_r$  is almost independent on whether perturbations are of short or long wavelength, whereas $r_d$ decreases for short and long wavelength perturbations with respect to the case $|k_x|\sim |k_z| \sim 1$. 
\section{Equilibrium profiles}
The classification developed in the previous section is general. One of the most interesting regimes is that of similar radial and vertical Richardson numbers and similar wavenumbers presented in section 2.3. We have seen that even for stratifications with $Ri_x>0$ and $Ri_z>0$ the disc can become unstable when $|{{L_S}_x}/{{L_S}_z}|$ is significantly away from 1. As a rule of thumb  for $|Ri_x|\sim| Ri_z|$ we would expect that $|{{L_S}_x}/{{L_S}_z}|\sim1$ and therefore no substantial growth of perturbation should occur.
This was indeed the case when we considered the vertically isothermal disc considered in Nelson, Gressel \& Umurhan (2013). No significant difference was found between $|{{L_S}_x}|$ and $|{{L_S}_z}|$ when $|Ri_x|\sim| Ri_z|$. As well for the cases $|Ri_z|>| Ri_x|$ and $|Ri_z|<| Ri_x|$ no growth was found for $Ri_x>0$ and $Ri_z>0$ in the regime of similar wavenumbers.
The vertically isentropic profile considered in Nelson, Gressel \& Umurhan (2013) is of scarce interest here since in that case $Ri_z=0$.

However, the above mentioned profiles are somewhat idealized since a realistic disc has both a temperature profile depending on the radial and vertical coordinates and a vertical Richardson number different from zero. In the following we consider two types of profiles in which the equilibrium temperature depends on both radial and vertical coordinates. The first profile is of central interest for the present study and pertains to a disc with $Ri_z>0$. 
The second profile describes a disc with $Ri_z<0$ and could be useful in the study of the interaction between the vertical convective and the vertical shear instabilities in an hydrodynamic disc.
\subsection{Profile with $Ri_z>0$}
We consider a profile where the density radial dependence at the midplane is identical to the one of the isothermal equilibrium studied in Nelson, Gressel \& Umurhan (2013), i.e.
\begin{equation}
\rho_{\rm mid}(R)=\rho_{0}\Big(\frac{R}{R_0}\Big)^p,
\label{eq1.40}
\end{equation}
where ${\rho}_0$ is the midplane density at the representative radius $R_0$.
Temperature, instead, acquires a $z$ dependence of the type 
\begin{equation}
T(R,z)=T_0\Big(\frac{R}{R_0}\Big)^q \Big(1+\frac{z^2}{H_0^2}\Big)^{1/2},
\label{eq1.41}
\end{equation}
where $T_0$ is the midplane temperature at $R_0$.
Assuming an ideal gas equation of state, equation~(\ref{eq1.41}) corresponds to
\begin{equation}
c_s^2(R,z)=c_0^2\Big(\frac{R}{R_0}\Big)^q \Big(1+\frac{z^2}{H_0^2}\Big)^{1/2},
\label{eq1.42}
\end{equation}
where $c_0^2=\gamma{\cal{R}}T_0/\mu$. In the above equations $H_0=c_0/\sqrt{GM/R_0^3}$, where $G$ is the gravitational constant and $M$ is the mass of the central object.

Solving the equilibrium equations
\begin{equation}
R{\Omega}^2-\frac{GMR}{(R^2+z^2)^{3/2}}=\frac{{\partial}_R P}{\rho},
\label{eq1.43a}
\end{equation}
\begin{equation}
-\frac{GMz}{(R^2+z^2)^{3/2}}=\frac{{\partial}_z P}{\rho},
\label{eq1.43b}
\end{equation}
subject to the equations of state (\ref{eq1.40}) and (\ref{eq1.42}) we obtain the equilibrium density and angular velocity profiles
\begin{eqnarray}
{\rho}_{\rm e}(R,z)=\rho_{0}\Big( \frac{R}{R_0}\Big)^p \Big(1+\frac{z^2}{H_0^2}\Big)^{-1/2}\nonumber\\  {\rm exp}\Big(\frac{\gamma GM}{c_0^2 ( {R}/{R_0})^q}\frac{H_0^2}{(R^2-H_0^2)}\Big[\frac{1}{R}-\frac{\sqrt{1+\frac{z^2}{H_0^2}}}{\sqrt{R^2+z^2}}\Big]\Big),
\label{eq1.44}
\end{eqnarray}
\begin{eqnarray}
\Omega_{\rm e}^2(R,z)=\Omega_K^2\Bigg\{\frac{1}{(1+z^2/R^2)^{3/2}}+\frac{H^2}{R^2}\frac{p+q}{\gamma}+\nonumber \\ \frac{H_0^2}{R^2-H_0^2} \sqrt{1+\frac{z^2}{H_0^2}}\Bigg[qR\Bigg(\frac{\sqrt{1+\frac{z^2}{H_0^2}}}{\sqrt{R^2+z^2}}-\frac{1}{R}\Bigg)+ \nonumber \\ 
\frac{1}{R^2-H_0^2}\Bigg(\sqrt{1+\frac{z^2}{H_0^2}}\frac{(3R^2+2z^2-H_0^2)}{(1+z^2/R^2)^{3/2}}- \nonumber\\ {(3R^2-H_0^2)}\Bigg)\Bigg]\Bigg\},
\label{eq1.45}
\end{eqnarray}
where $\Omega_K^2={GM}/{R^3}$ is the Keplerian angular velocity and $H=c_s/\Omega_K$ a local scale height depending on $R$ and $z$.

To have a notation consistent with the equilibrium discussed in this section we rename the radial Richardson number and length scales in terms of the radius $R$, i. e.
\begin{equation}
\frac{1}{L_{P_R}}\equiv\frac{\partial_{R}P_{\rm e}}{\gamma P_{\rm e}}=\frac{1}{{L_{{\rho}_R}}}+\frac{1}{L_{S_R}}\equiv\frac{\partial_{R}{\rho}_{\rm e}}{ {\rho}_{\rm e}}+\frac{\partial_{R}S_{\rm e}}{\gamma S_{\rm e}},
\label{eq1.18b}
\end{equation}
\begin{equation}
{Ri}_R \equiv-\frac{H^2}{L_{S_R}L_{P_R}},
\label{eq1.30b}
\end{equation}
keeping in mind that $Ri_x=Ri_R$.

\begin{figure}
\epsfxsize=3.0in
\epsfysize=2.0in
	\begin{center}
		\includegraphics[scale=1]{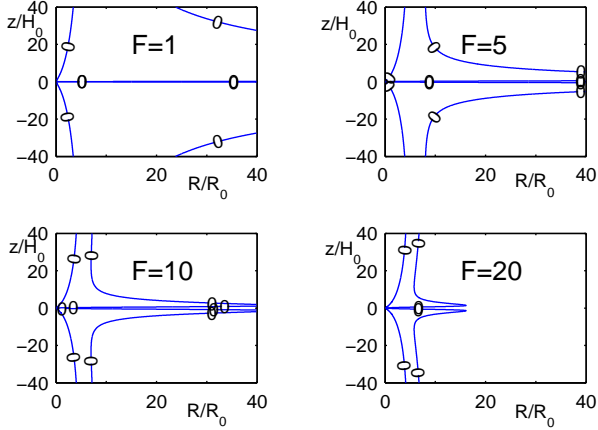}

	\end{center}
	\caption{Contours of $Ri_z-FRi_R=0$ relative to the cases $F=1, 5, 10, 20$ for profile~(\ref{eq1.41}) ($p=-1.5$ and $q=-0.3$).}
\label{axifig:0}
\end{figure}

In this section we will study contours of the type $Ri_z=F Ri_R$, where $F$ is a real number, for different values of $F$. In Fig.~\ref{axifig:0} we  show the shape of such contours and how they change with increasing $F$.
In general a contour is composed of lines lying in the proximity of the midplane, from now on denoted as "internal", and lines positioned further out vertically, characterized hereafter as "external". When $F$ is increased the contours tend to shrink both radially and vertically. In the following we will mainly investigate $Ri_z=F Ri_R$ in the domain $(R/R_0,z/H_0)\in[0,20]\times[-20,20]$.

We start our analysis with the case $F=1$. We need now to determine the set of points $(R,z)$ solving the equation $Ri_R=Ri_z$ and there find the value of the ratio ${{L_S}_R}/{{L_S}_z}$. From the equilibrium equations (\ref{eq1.44}) and (\ref{eq1.45}) it is straightforward to find the analytical expression of these quantities. They are reported in Appendix. Fig.~\ref{axifig:2} shows the graph of $Ri_z$.\\
We set $p=-1.5$ and studied the cases $q=-0.1$, $q=-0.3$, $q=-0.5$ for a disc of aspect ratio ${\cal A}=H_0/R_0=0.1$ and $\gamma=1.4$. In all these cases the physically relevant part of $Ri_R=Ri_z$ occurs in the vicinity of the midplane as shown in Fig.~\ref{axifig:1} for $q=-0.3$.  Fig.~\ref{axifig:1} is a close-up of the upper left contour of Fig.~\ref{axifig:0}.
To estimate the vertical velocity shear $A_z$ we make use of eq.~(\ref{eq1.5C}) obtaining the expression
\begin{equation}
A_z \approx \frac{z}{2R}\Big[\Big(\frac{R}{R_0}\Big)^q\Big(1+\frac{z^2}{H_0^2}\Big)^{-1/2}\frac{p+q}{\gamma}+q\Big],
\end{equation}
which for the contour of Fig.~\ref{axifig:1} gives a reference value $|A_z|\approx0.001$.
\begin{figure}
\epsfxsize=3.0in
\epsfysize=2.0in
	\begin{center}
		\includegraphics[scale=1]{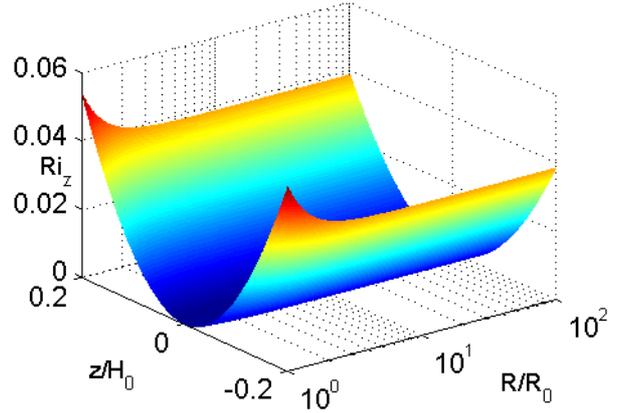}

	\end{center}
	\caption{$Ri_z$ for profile~(\ref{eq1.41}) ($p=-1.5$ and $q=-0.3$).}
\label{axifig:2}
\end{figure}
\begin{figure}
\epsfxsize=3.0in
\epsfysize=2.0in
	\begin{center}
		\includegraphics[scale=1]{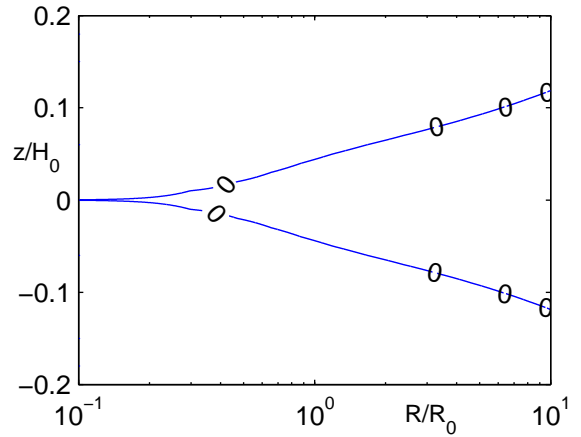}

	\end{center}
	\caption{Close-up of the contour $Ri_R-Ri_z=0$ near the midplane for profile~(\ref{eq1.41}) ($p=-1.5$ and $q=-0.3$).}
\label{axifig:1}
\end{figure}

Along the contour, $Ri_z$ and ${{L_S}_R}/{{L_S}_z}$ in general change. To these different values correspond different growth rates. In Table 1 we list the growth rates (last column) at $R=10 R_0$ relative to the case of thermal diffusion. Similarly Table 2 reports growth rates in the case of thermal relaxation. Maximum growth rates (about $0.25\Omega$) pertain to values of $q>p$ and closest to 0 (here we considered always $q<0$). We notice that at locations symmetric to the ones reported in Tables 1 and 2 with respect to the midplane, the ratio ${{L_S}_R}/{{L_S}_z}$ assumes opposite values and instability occurs with the same growth rate when ${{L_S}_R}/{{L_S}_z}$ and $k_x k_z$ have the same sign. For instance, in the case of $q=-0.1$ for $(R,z)=(10R_0,-0.18H_0)$ we have ${{L_S}_R}/{{L_S}_z}=-50$ and instability occurs with $s_d=0.23$ for $k_x=-k_z=\pm1$.
\begin{table}
\caption {Growth rates $s_d$ for profile~(\ref{eq1.41}) along the contour $Ri_R=Ri_z$ in the case of thermal diffusion for $k_x=k_z=1$, aspect ratio ${\cal A}=0.1$ and vertical shear $A_z=0.001$. $Ri$ is the common value of $Ri_R$ and $Ri_z$ at the $R$ and $z$ specified}  
\label{tab:1} 
\begin{center}
    \begin{tabular}{ | p{0.5cm} | p{0.5cm} | p{0.5cm} |  p{0.5cm} | p{0.5cm} | p{0.5cm} | p{0.5cm} | p{0.5cm} |}
    \hline
   \hspace{0.2cm} $p$ & \hspace{0.2cm}$q$ & $\frac{R}{R_0}$& $\frac{z}{H_0}$ &$Ri$& $\frac{{L_S}_R}{{L_S}_z}$&   $\eta$ & $s_d$  \\ \hline
    $-1.5$  & $-0.1$ &  $10$ & $0.18$ &$0.031$ & $50$ & $0.3$  & $0.23$ \\ \hline
    $-1.5$  & $-0.3$ &  $10$ & $0.12$ &$0.014$ & $55$ & $0.3$ & $0.14$ \\ \hline
    $-1.5$  & $-0.5$ &  $10$ & $0.055$ &$0.003$ & $77$ & $0.3$ & $0.041$ \\ \hline
    \end{tabular}
\end{center}
\end{table}
\begin{table}
\caption {Growth rates $s_r$ for profile~(\ref{eq1.41}) along the contour $Ri_R=Ri_z$ in the case of thermal relaxation for $k_x=k_z=1$, aspect ratio ${\cal A}=0.1$  and vertical shear $A_z=0.001$. $Ri$ is the common value of $Ri_R$ and $Ri_z$ at the $R$ and $z$ specified}  
\label{tab:2} 
\begin{center}
    \begin{tabular}{ | p{0.5cm} | p{0.5cm} | p{0.5cm} |  p{0.5cm} | p{0.5cm} | p{0.5cm} | p{0.5cm} | p{0.5cm} |}
    \hline
   \hspace{0.2cm} $p$ & \hspace{0.2cm}$q$ & $\frac{R}{R_0}$& $\frac{z}{H_0}$ &$Ri$& $\frac{{L_S}_R}{{L_S}_z}$&   ${\bar{\chi}}_r$ & $s_r$  \\ \hline
    $-1.5$  & $-0.1$ &  $10$ & $0.18$ &$0.031$ & $50$ & $0.3$  & $0.27$ \\ \hline
    $-1.5$  & $-0.3$ &  $10$ & $0.12$ &$0.014$ & $55$ & $0.3$ & $0.14$ \\ \hline
    $-1.5$  & $-0.5$ &  $10$ & $0.055$ &$0.003$ & $77$ & $0.3$ & $0.037$ \\ \hline
    \end{tabular}
\end{center}
\end{table}
We concentrated then our attention on the representative case $q=-0.3$. In Table 3 we present the growth rates at various locations along the contour $Ri_R=Ri_z$ for the case of thermal diffusion. They vary greatly, increasing with $R$. We notice as well that due to a scaling symmetry of eq.~(\ref{eq1.35C}) the same results reported in Table 3 hold if we increase the order of $k_x$ and $k_z$ and decrease the order of $\eta$ of twice the order of the wavenumbers. For example if we consider perturbations with $k_x=k_z=10$ the same growth rates of Table 3 are found when $\eta=0.003$.\\
Growth rates for the thermally relaxed case are slightly higher than the ones of Table 3. As previously mentioned, the main difference between thermal diffusion and thermal relaxation consists in the fact that $s_r$ is wavenumber independent. In the case of thermal diffusion the growth rate $s_d$ decreases for $|k_x|,|k_z|\gg1$ or $ |k_x|,|k_z|\ll1$. We notice that all growth rates reported were obtained by numerically solving the exact equations and not from the approximations~(\ref{eq1.35CsimplifiedGrowthratediff}) and~(\ref{eq1.35CsimplifiedGrowthraterelax}).\\
\begin{table}
\caption {Growth rates $s_d$ for profile~(\ref{eq1.41}) along the contour $Ri_R=Ri_z$ in the case of thermal diffusion for $p=-1.5$, $q=-0.3$, $k_x=k_z=1$, ${\cal A}=0.1$  and $A_z=0.001$. $Ri$ is the common value of $Ri_R$ and $Ri_z$ at the $R$ and $z$ specified. "ND" stays for not definite.}  
\label{tab:3} 
\begin{center}
    \begin{tabular}{ | p{0.5cm} |  p{0.5cm} | p{0.5cm} | p{0.5cm} | p{0.5cm} | p{0.5cm} | p{0.5cm} |}
    \hline
    $ \hspace{-0.1cm}\frac{R}{R_0}$& $\frac{z}{H_0}$ &$Ri$& $\frac{{L_S}_R}{{L_S}_z}$&   $\eta$ & $s_d$ & $W_{xy}$ \\ \hline
     $2$ & $0.06$ &$0.0038$ & \hspace{0.3cm}$6$ & $0.3$  & $0.004$ & $ND$ \\ \hline
     $5$ & $0.085$ &$0.0072$ & \hspace{0.1cm} $20$ & $0.3$  & $0.025$ & $ND$ \\ \hline
     $10$ & $0.12$ &$0.014$ &  \hspace{0.2cm}$55$ & $0.3$ & $0.14$ & $ND$\\ \hline
     $15$ & $0.13$ &$0.016$ &  \hspace{0.2cm}$90$ & $0.3$ & $0.24$ & $ND$\\ \hline
     $30$ & $0.18$ &$0.031$ &  \hspace{0.05cm}$244$ & $0.3$ & $1.45$ & $-$\\ \hline
    \end{tabular}
\end{center}
\end{table}
An important point is the direction of the angular momentum transport. In all cases the sign of the Reynolds stress $W_{xy} \equiv ({v_x}{v_y})/ {v}^2$, where ${v}^2={v_x}^2+{v_y}^2+{v_z}^2$,  is not definite apart from the last line in Table 3 where the transport was observed to be negative.\\
By increasing $q$ the instability is more powerful. Here, as in Klahr and Hubbard (2014), we observe maximum growth when ${\bar{\chi}}_r \sim 1$.

To determine whether the instability conditions~(\ref{eq1.35instconv3}) and~(\ref{eq1.35instconv5}) are met for the equilibrium~(\ref{eq1.41}), we consider the equaton $Ri_z=F Ri_R$ and examine the cases $F>1$ (i.e. $Ri_z > Ri_R$) and $F<1$ (i.e. $Ri_z < Ri_R$).\\
 We found that for $F<1$  the internal part of the contour approaches the midplane, where $Ri_R$ and $Ri_z$ are too small and ${{L_S}_R}/{{L_S}_z}$ too close to 1 for significant growth to occur. The external part moves instead further outward in regions not physically relevant.\\
More interesting is the case $F>1$, in which the internal lines of the contour move away from the midplane and $Ri_R$, $Ri_z$ and ${{L_S}_R}/{{L_S}_z}$ increase substantially. The external lines instead move closer to the midplane. All such contours are highly unstable. In Fig. 4 we present a close up of the internal lines of the contour pertaining to $F=10$ with the corresponding growth rates given in Table 4, where as well can be found the ones relative to the cases $F=3, 5$. In Table 5 we present the growth rates relative to the cases $F=5, 10, 20$ at representative locations on the external lines of the contours. We notice that for all contours with $F>1$ we have $Ri_R({{L_S}_R}/{{L_S}_z})>Ri_z$. From this the instability stems, in agreement with condition~(\ref{eq1.35instconv5}).\\
\begin{figure}
\epsfxsize=3.0in
\epsfysize=2.0in
	\begin{center}
		\includegraphics[scale=1]{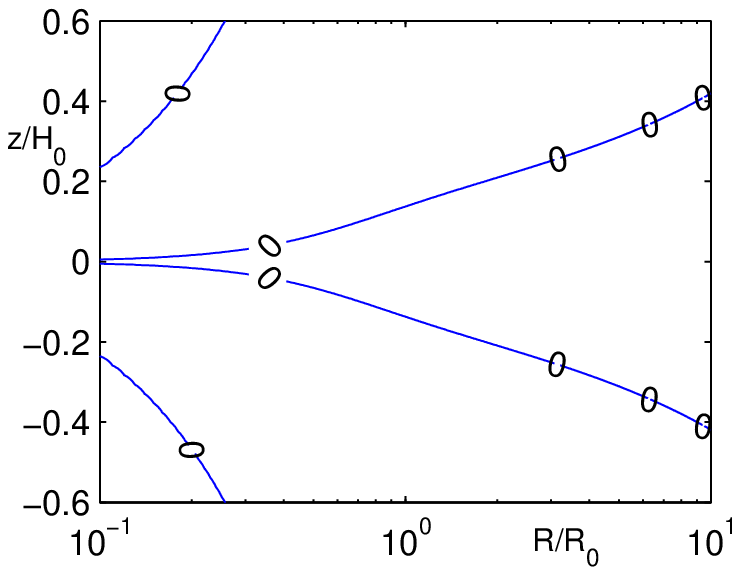}

	\end{center}
	\caption{Close-up of the contour $Ri_z-10Ri_R=0$ in the proximity of the midplane for profile~(\ref{eq1.41}) ($p=-1.5$ and $q=-0.3$).}
\label{axifig:3}
\end{figure}
Again it is important to ascertain the sign of the angular momentum transport, which turns out to be not definite for slower growth rates and negative for faster ones.\\
\begin{table}
\caption {Growth rates $s_d$ for equilibrium~(\ref{eq1.41}) along the internal lines of contour $Ri_z=F Ri_R$ in the case of thermal diffusion for $p=-1.5$, $q=-0.3$, $k_x=k_z=1$, ${\cal A}=0.1$  and $A_z=0.001$. "ND" stays for not definite.}  
\label{tab:4} 
\begin{center}
    \begin{tabular}{ | p{0.5cm} |  p{0.5cm} | p{0.5cm} | p{0.5cm} | p{0.5cm} | p{0.5cm} | p{0.5cm} | p{0.5cm} |}
    \hline
    $ F $&$ \hspace{-0.1cm}\frac{R}{R_0}$& $\frac{z}{H_0}$ &$Ri_z$& $\frac{{L_S}_R}{{L_S}_z}$&   $\eta$ & $s_d$ & $W_{xy}$ \\ \hline
    $3$ & $2$ & $0.11$ &$0.013$ & \hspace{0.1cm}$11$ & $0.3$  & $0.005$ & $ND$\\ \hline
     $3$ &$5$ & $0.16$ &$0.025$ & \hspace{0.1cm}$37$ & $0.3$  & $0.046$ & $ND$\\ \hline
     $3$ &$10$ & $0.21$ &$0.042$ & \hspace{0.1cm}$94$ & $0.3$ & $0.207$ & $ND$\\ \hline
     $3$ &$15$ & $0.25$ &$0.056$ & \hspace{0.05cm}$162$ & $0.3$ & $0.414$ & $ND$\\ \hline
     $3$ &$20$ & $0.27$ &$0.068$ & \hspace{0.05cm}$235$ & $0.3$ & $1.081$ & $-$\\ \hline
     $5$ &$2$ & $0.15$ &$0.023$ & \hspace{0.1cm}$15$ & $0.3$  & $0.008$ & $ND$\\ \hline
     $5$ &$5$ & $0.22$ &$0.046$ & \hspace{0.1cm}$49$ & $0.3$  & $0.069$ & $ND$\\ \hline
     $5$ &$10$ & $0.27$ &$0.068$ &  \hspace{0.05cm}$118$ & $0.3$ & $0.25$ & $ND$\\ \hline
     $5$ &$15$ & $0.33$ &$0.098$ &  \hspace{0.05cm}$208$ & $0.3$ & $0.805$ & $-$\\ \hline
     $5$ &$20$ & $0.37$ &$0.12$ &  \hspace{0.05cm}$304$ & $0.3$ & $1.38$ & $-$\\ \hline
     $10$ &$2$ & $0.21$ &$0.045$ & \hspace{0.1cm}$20$ & $0.3$  & $0.009$ & $ND$\\ \hline
     $10$ &$5$ & $0.31$ &$0.088$ & \hspace{0.1cm}$67$ & $0.3$  & $0.092$ & $ND$\\ \hline
     $10$ &$10$ & $0.42$ &$0.15$ &  \hspace{0.05cm}$167$ & $0.3$ & $0.345$ & $ND$\\ \hline
     $10$ &$15$ & $0.5$ &$0.20$ &  \hspace{0.05cm}$280$ & $0.3$ & $1.081$ & $-$\\ \hline
     $10$ &$20$ & $0.6$ &$0.26$ &  \hspace{0.05cm}$412$ & $0.3$ & $1.82$ & $-$\\ \hline
    \end{tabular}
\end{center}
\end{table}
\begin{table}
\caption {Growth rates $s_d$ for equilibrium~(\ref{eq1.41}) along external lines of the contour $Ri_z=F Ri_R$ in the case of thermal diffusion for $p=-1.5$, $q=-0.3$, $k_x=k_z=1$, ${\cal A}=0.1$  and $A_z=0.001$. "ND" stays for not definite.}  
\label{tab:4} 
\begin{center}
    \begin{tabular}{ | p{0.5cm} |  p{0.5cm} | p{0.5cm} | p{0.5cm} | p{0.7cm} | p{0.5cm} | p{0.5cm} | p{0.5cm} |}
    \hline
    $ F $&$ \hspace{-0.1cm}\frac{R}{R_0}$& $\frac{z}{H_0}$ &$Ri_z$& $\frac{{L_S}_R}{{L_S}_z}$&   $\eta$ & $s_d$ & $W_{xy}$ \\ \hline
     $5$ &$3$ & $20$ &$0.71$ & \hspace{0.0cm}$-3.3$ & $0.3$  & $decay$ & $ND$\\ \hline
     $5$ &$10$ & $18$ &$0.96$ &  \hspace{0.05cm}$31$ & $0.3$ & $0.97$ & $-$\\ \hline
     $5$ &$20$ & $8.5$ &$0.98$ &  \hspace{0.05cm}$110$ & $0.3$ & $2.78$ & $-$\\ \hline
     $10$ &$7$ & $20$ &$0.92$ & \hspace{0.1cm}$32$ & $0.3$  & $0.3$ & $ND$\\ \hline
     $10$ &$10$ & $7.5$ &$0.98$ &  \hspace{0.05cm}$66$ & $0.3$ & $1.08$ & $-$\\ \hline
     $10$ &$20$ & $4.2$ &$0.94$ &  \hspace{0.05cm}$211$ & $0.3$ & $2.64$ & $-$\\ \hline
     $20$ &$7$ & $5$ &$0.96$ & \hspace{0.1cm}$71$ & $0.3$  & $0.34$ & $ND$\\ \hline
     $20$ &$10$ & $2.8$ &$0.88$ &  \hspace{0.05cm}$151$ & $0.3$ & $1.15$ & $-$\\ \hline
     $20$ &$16$ & $1.6$ &$0.72$ &  \hspace{0.05cm}$336$ & $0.3$ & $1.93$ & $-$\\ \hline
    \end{tabular}
\end{center}
\end{table}

The linear theory therefore predicts inward transport of angular momentum in the external layers of the disc or for larger radii at robust rates and indefinite sign in the interior.
\subsection{Profile with $Ri_z<0$}
\begin{figure}
\epsfxsize=3.0in
\epsfysize=2.0in
	\begin{center}
		\includegraphics[scale=1]{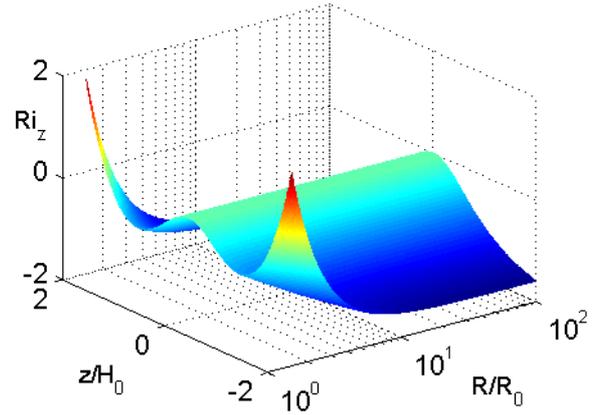}

	\end{center}
	\caption{$Ri_z$ for profile~(\ref{eq1.41B}) for $p=-1.5$ and $q=-1$.}
\label{axifig:5}
\end{figure}

We consider here an equilibrium with midplane density identical to the one discussed in the previous section (equation~(\ref{eq1.40})) and a temperature profile of the type
\begin{equation}
T(R,z)=T_0\Big(\frac{R}{R_0}\Big)^q \Big(1+\frac{z^2}{H_0^2}\Big)^{-1},
\label{eq1.41B}
\end{equation}
which corresponds for an ideal gas to 
\begin{equation}
c_s^2(R,z)=c_0^2\Big(\frac{R}{R_0}\Big)^q \Big(1+\frac{z^2}{H_0^2}\Big)^{-1}.
\label{eq1.42B}
\end{equation}
Solving the equilibrium equations~(\ref{eq1.43a}) and~(\ref{eq1.43b}) we have
\begin{eqnarray}
\rho(R,z)=\rho_{0}\Big( \frac{R}{R_0}\Big)^p \Big(1+\frac{z^2}{H_0^2}\Big)\nonumber\\  {\rm exp}\Big(\frac{\gamma GM}{c_0^2 ( {R}/{R_0})^q}\frac{1}{H_0^2}\Big[\frac{2R^2-H_0^2}{R}-\nonumber\\ \frac{2R^2+z^2-H_0^2}{\sqrt{R^2+z^2}}\Big]\Big),
\label{eq1.44B}
\end{eqnarray}
\begin{eqnarray}
\Omega^2(R,z)=\Omega_K^2\Bigg\{\frac{1}{(1+{z^2}/{R^2})^{3/2}}+\nonumber \\ \frac{H^2}{R^2}\frac{p+q}{\gamma}+ \frac{1}{H_0^2} {\Big(1+\frac{z^2}{H_0^2}\Big)^{-1}} \nonumber \\ \Bigg[q\Bigg(\frac{2R^2+z^2-H_0^2}{(1+{z^2}/{R^2})^{1/2}}-{(2R^2-H_0^2)}\Bigg)- \nonumber \\ 
\Bigg(\frac{(2R^2+3z^2+H_0^2)}{(1+{z^2}/{R^2})^{3/2}}- {(2R^2+H_0^2)}\Bigg)\Bigg]\Bigg\}.
\label{eq1.45B}
\end{eqnarray}
This profile could be useful to study the interaction of vertical shear and vertical convective instabilities. In Fig.~\ref{axifig:5} we show the graph of the vertical Richardson number corresponding to this equilibrium.

\section{Summary and Discussion}
We described the convective stability of a fully stratified disc. We have considered the sector of the dispersion relation derived by Urpin (2003) pertaining to the regime $|Ri_z|\geq |Ri_x|$ and $|k_x|\sim |k_z|$ finding that even in the case of $Ri_x>0$ and $Ri_z>0$ the disc can be destabilized for values of the ratio $|{{L_S}_R}/{{L_S}_z}|$ significantly away from 1. We presented as well an equilibrium profile where this condition can be realized. The instability is very strong in the outer layers or for larger radii with inward transport of angular momentum occurring there, while in internal layers growth is more contained with indefinite sign of $W_{xy}$.

The overall picture arising is one where an astrophysical disc is potentially teeming with instabilities of different origins. Perturbations with $|k_x| \gg |k_z|$ grow due to the vertical shear instability (Urpin 2003; Nelson, Gressel \& Umurhan 2013) or the vertical convective instability ($Ri_z<0$),  if present, or a combination of the two (Volponi 2014). On the other side the ones with $|k_z| \gg |k_x|$ are subject to growth in zones of the discs where the condition $Ri_x<0$ (Urpin 2003; Klahr \& Hubbard 2014)  pertains. For the  intermediate regime  $|k_x|\sim |k_z|$ the disc can be destabilized owing to the hybrid Richardson number $Ri_{xz}={\rm sgn}(k_x k_z)[({{L_S}_x}/{{L_S}_z})Ri_x+({{L_S}_z}/{{L_S}_x})Ri_z]$.

One important point concerns the sign of the angular momentum transport associated to linear perturbations. Linear theory predicts outward transport for the vertical shear instability, inward transport for vertical convection and no definite sign for the convective overstability. For the hybrid convective instability here presented we mentioned above that the sign is either negative or non definite. In the studies of Klahr \& Hubbard (2014) and Lyra (2014) the convective overstability was identified as the triggering mechanism for the subcritical baroclinic instability. In those studies in the nonlinear regime the $\alpha$ parameter was found to be of order $10^{-3}$ and inducing an outward transport of angular momentum. In this sense it appears that an inherently nonlinear mechanism is at work in determining the outward direction of the transport, since the sign of $W_{xy}$ is not definite for the convective overstability. The Reynolds stress of linear perturbations, $W_{xy}$, is usually a good indicator of the direction of the angular momentum transport in the nonlinear regimes. This happens for the magnetorotational, vertical shear and vertical convective instabilities for which linear and nonlinear Reynolds stresses have the same sign, positive for the first two instabilities and negative for the third one.

The ratio ${{L_S}_x}/{{L_S}_z}$ plays a central role in the instability mechanism under consideration. In the equilibrium of section 3.1 its scaling for $R>R_0$ is given with good approximation by 
\begin{equation}
\frac{{{L_S}_x}}{{{L_S}_z}}\sim\frac{\gamma {\cal{A}}^{-1}}{q+p(1-\gamma)}\frac{R}{R_0}\frac{z/H_0}{{1+(z/H_0)^2}}.
\end{equation}
At height $z\neq0$, ${{L_S}_x}/{{L_S}_z}$ increases linearly with $R$, whereas it becomes zero at $z=0$. We expect, therefore, ${{L_S}_x} \gg {{L_S}_z} $ at large radii and ${{L_S}_z} \gg {{L_S}_x} $ at the midplane. As well, when $z\rightarrow 0$ we obtain the limits ${{L_S}_z} \sim 1/z \rightarrow \infty$ and $Ri_z  \sim z^2 \rightarrow 0$. It follows that the midplane is a region where the condition $Ri_{xz}=0$ holds.

We conclude by summarizing the three ingredients necessary for the onset of the instability here investigated. The first is the presence in the disc's domain of surfaces $Ri_z=F Ri_R$. The second is a not too small value of the Richardson numbers on these surfaces. Finally a ratio ${{L_S}_R}/{{L_S}_z}$ significantly away from 1 is required there. Roughly, growth occurs when either $Ri_R({{L_S}_R}/{{L_S}_z})$ or $Ri_z({{L_S}_z}/{{L_S}_R})$ is of order one. 
The equilibrium here discussed fullfills the three conditions above specified. However for a similar equilibrium temperature, i.e. $T(R,z)=T_0({R}/{R_0})^q [1+(z^2/{H_0^2})(R/R_0)^{-(q+3)}]^{1/2}$, we noticed that both $Ri_R({{L_S}_R}/{{L_S}_z})$ and $Ri_z({{L_S}_z}/{{L_S}_R})$ are always much smaller than 1 and therefore growth, when present, is weak (we obtained characteristic growth rates of about $0.005\Omega$ for $p=-1.5$ and $q=-0.5$). A sistematic analysis of realistic equilibria which are susceptible to destabilization owing to the hybrid Richardson number is a matter which is left open by the present study and which will be worth investigating in the future.
\section*{Acknowledgements}
The author is grateful to the reviewer of this paper for suggestions which improved the original manuscript.

\appendix
\section*{Appendix}
\renewcommand{\theequation}{\Alph{section}A.\arabic{equation}}
Here we present the formulas for radial and vertical Richardson numbers and length scales for the equilibrium of section 3.1.
We are considering sound speed with dependence of the type
\begin{equation}
c_s^2(R,z)=c_0^2 f(R) g_1(z),
\label{eqA1}
\end{equation}
where 
\begin{equation}
f(R)=\Big(\frac{R}{R_0}\Big)^q \hspace{0.3cm} {\rm and} \hspace{0.3cm} g_1(z)=\Big(1+\frac{z^2}{H_0^2}\Big)^{1/2}.
\label{eqA2}
\end{equation}
Recalling that $S=P{\rho}^{-\gamma}$ and that $c_s^2=\gamma P/\rho$ we have
\begin{equation}
\frac{1}{{L_S}_R}=\frac{1-\gamma}{\gamma}\frac{1}{{L_{\rho}}_R}+\frac{1}{\gamma f}\frac{{\rm d}f}{{\rm d} R},
\label{eqA3}
\end{equation}
\begin{equation}
\frac{1}{{L_S}_z}=\frac{1-\gamma}{\gamma}\frac{1}{{L_{\rho}}_z}+\frac{1}{\gamma g_1}\frac{{\rm d}g_1}{{\rm d} z}.
\label{eqA4}
\end{equation}
For the density profile of equation (\ref{eq1.44}), considering that
\begin{equation}
\frac{1}{f}\frac{{\rm d}f}{{\rm d} R}=\frac{q}{R} \hspace{0.3cm} {\rm and} \hspace{0.3cm} \frac{1}{g_1}\frac{{\rm d}g_1}{{\rm d} z}=\frac{z}{z^2+H_0^2},
\end{equation}
equations (\ref{eqA3}) and (\ref{eqA4}) become
\begin{eqnarray}
\frac{1}{{L_S}_R}=\frac{1-\gamma}{\gamma}\Bigg[\frac{p}{R}-\frac{q}{R}\frac{\gamma GM}{c_0^2(R/R_0)^q}\frac{H_0^2}{R^2-H_0^2} \nonumber \\ 
\Big(\frac{1}{R}-\frac{\sqrt{1+\frac{z^2}{H_0^2}}}{\sqrt{R^2+z^2}}\Big)+
\frac{\gamma GM}{c_0^2(R/R_0)^q}\frac{H_0^2}{(R^2-H_0^2)^2}\nonumber \\ \Big(\frac{R(3R^2+2z^2-H_0^2)\sqrt{1+\frac{z^2}{H_0^2}}}{(R^2+z^2)^{3/2}}- \nonumber \\ \frac{3R^2-H_0^2}{R^2}\Big)\Bigg]+\frac{q}{\gamma R},
\label{eqA5}
\end{eqnarray}
\begin{equation}
\frac{1}{{L_S}_z}=\frac{z}{z^2+H_0^2}+\frac{(\gamma-1)GM}{c_s^2}\frac{z}{(R^2+z^2)^{3/2}}.
\label{eqA7}
\end{equation}
The radial and vertical Richardson numbers are easily expressed as 
\begin{eqnarray}
Ri_R=\frac{1}{\gamma-1}\frac{H^2}{{L_{S}}_R}\Big(\frac{1}{{{L_S}_R}}-\frac{{\partial}_R f}{f}\Big), \nonumber \\
Ri_z=\frac{1}{\gamma-1}\frac{H^2}{{L_{S}}_z}\Big(\frac{1}{{{L_S}_z}}-\frac{{\partial}_z g}{g}\Big).
\label{eqA8}
\end{eqnarray}
We can as well express all the above quantities by means of the nondimensional variables $\sigma=R/R_0$ and $\zeta=z/H_0$ obtaining
\begin{eqnarray}
\frac{R_0}{{L_S}_R}=\frac{q+p(1-\gamma)}{\gamma \sigma}+\frac{1-\gamma}{{\sigma}^{q+2}(\sigma^2-{\cal A}^2)}\nonumber \\ \Bigg[q \Big(\frac{\sqrt{1+{\zeta}^2}}{\sqrt{1+{\cal A}^2\frac{{\zeta}^2}{{\sigma}^2}}}-1\Big)+
\frac{1}{({\sigma}^2-{\cal A}^2)}\nonumber \\ \Big(\frac{3{\sigma}^2+(2{\zeta}^2-1){\cal A}^2}{({1+{\cal A}^2\frac{{\zeta}^2}{{\sigma}^2}})^{3/2}}\sqrt{1+{\zeta}^2}- ({3{\sigma}^2-{\cal A}^2})\Big)\Bigg],
\label{eqA9}
\end{eqnarray}
\begin{eqnarray}
\frac{R_0}{{L_S}_z}={\cal A}^{-1}\Big(\frac{\zeta}{1+{\zeta}^2}+ \nonumber \\ \frac{\zeta}{({1+{\cal A}^2\frac{{\zeta}^2}{{\sigma}^2}})^{3/2}}\frac{\gamma-1}{{\sigma}^{q+3}{\sqrt{1+{\zeta}^2}}}\Big),
\label{eqA10}
\end{eqnarray}
\begin{eqnarray}
Ri_R=\frac{{\cal A}^2{\sigma^{q+3}\sqrt{1+{\zeta}^2}}}{\gamma-1}\Big(\frac{R^2_0}{{{L^2_S}_R}}-\frac{R_0}{{L_{S}}_R}\frac{q}{\sigma}\Big), \nonumber \\
Ri_z=\frac{{\cal A}{\sigma^{q+3}\sqrt{1+{\zeta}^2}}}{\gamma-1}\Big({\cal A}\frac{R^2_0}{{{L^2_S}_z}}-\frac{R_0}{{L_{S}}_z}\frac{\zeta}{1+{\zeta}^2}\Big).
\label{eqA11}
\end{eqnarray}

\end{document}